\documentclass[pdftex]{elsart}
\usepackage[]{epsfig}
\usepackage{amsmath}
\usepackage{pstricks}
\begin{document}
 
\begin{frontmatter}

	\title{A particle-like description of Planckian  black holes}
	\author{Euro Spallucci\thanksref{infn}}
\thanks[infn]{e-mail address: spallucci@ts.infn.it }
\address{Dipartimento di Fisica Teorica, Universit\`a di Trieste
and INFN, Sezione di Trieste, Italy}
 
\author{Anais Smailagic\thanksref{infn2}}
\thanks[infn2]{e-mail address: anais@ts.infn.it }
\address{Dipartimento di Fisica Teorica, Universit\`a di Trieste
and INFN, Sezione di Trieste, Italy}
        
	\begin{abstract}
	In this paper we abandon the idea that even a ``quantum'' black hole, of Planck size,
        can still be described as a \emph{classical}, more or less complicated, geometry.
        Rather, we consider a genuine quantum mechanical approach where a Planckian black hole is, by all means, just another
        ``particle'', even if with a distinguishing  property: its wavelength increases with the energy.
       The horizon dynamics is equivalently described in terms of a particle  moving in gravitational
         potential derived from the horizon equation itself in a self-consistent manner.
        The particle turning-points match the radius of the inner and outer horizons of a charged black hole. 
        This classical model pave the way  towards the wave equation for a truly quantum black hole.
        We compute the exact form of the wave function and determine the energy spectrum. Finally, we describe 
        the classical limit in which the quantum picture correctly approaches the classical geometric formulation. 
       We find that the quantum-to-classical transition occurs far above the Planck scale. 
        \end{abstract}
	\end{frontmatter}
	
\section{Introduction}

Since the introduction of the concept of radiating , ``mini'', black holes by Hawking \cite{Hawking:1975iha}, there has been an increasing interest for
black holes (BHs) which are not produced by the gravitational collapse of stellar size masses, but for those that have linear size comparable, 
or even smaller, than an elementary particle.  Despite the ``\emph{abyssal}'' difference in size and mass between a galactic center BH of
billion solar masses,  and a theoretical micro quantum BH smaller than an atomic nucleus, the formal description of such very different
objects remains the same. In both cases one has in mind classical solutions of Einstein equations, i.e. a classical geometrical description,
with the only difference that cosmic objects interact with classical matter, while micro BHs interacts with quantum particles.\\
This state of mind has led to various models of quantum BHs in which the ``quantum'' nature is simulated through non-trivial geometrical
and topological distortions, e.g. ``large'' or ``warped'' extra-dimensions. In this framework, the restriction to look for ``imprints''
of mini-BHs existence in  the early universe only, can be avoided by opening the exciting possibility to study them in the lab through
high energy  particle  collisions.\\
 The standard approach to ``quantum'' BHs is  motivated by the generally accepted idea that
true quantum gravity effects will manifest themselves only near the Planck energy scale. Therefore, BHs much smaller than a proton,
can still be considered ``\emph{classical}'' objects, as long as their size is large with respect the Planck length $l_P= 10^{-33}\, cm $.
The main shortcoming of this ``\emph{scale downgrading}'' approach is that it breaks down just near the Planck scale where it is supposed
that these objects should be produced!\\
A clear example of this failure, is that the final stage of the BH thermal decay cannot be defined except for BHs admitting an 
\emph{extremal} configuration. Even in this case, the third law of thermodynamics seems to be violated, since the temperature is zero,
but the entropy is given by the \emph{non-vanishing} area of the degenerate horizon. Last but not the least, 
the statistical description in terms  of micro-states remains confined to a limited  number of special super-symmetric models.\\

Against this background, we would like to propose the idea of ``\emph{energy scale upgrade}'' in the sense that we start from elementary
particles  below  the Planck scale and gradually approach the Planck phase from below. This line of reasoning is inspired by  the
UV self-complete quantum gravity program introduced in \cite{Dvali:2010bf,Dvali:2010jz}. In this picture hadronic collisions at Planckian 
energy
\cite{Rizzo:2002kb,Chamblin:2002ad,Mocioiu:2003gi,Lonnblad:2005ah,Rizzo:2006di,Calmet:2008dg,Najafabadi:2008xv,Chamblin:2008ec,Erkoca:2009kg}, 
\cite{Kiritsis:2011qv,Spallucci:2011rn,Mureika:2011hg,Spallucci:2012xi,Casadio:2013uga,Nicolini:2013ega,Alberghi:2013hca} can result 
in the production of ``non-geometrical'' BHs 
described as  Bose-Einstein graviton condensates\cite{Dvali:2011aa,Dvali:2012gb,Dvali:2012en,Dvali:2013vxa,Casadio:2013hja}.
\\ 
Stimulated by the hope that this new scenario can cure previously described limitations of the ``scale downgrading'' approach, and give
new insight into the quantum nature of BHs, we build a quantum model ``from scratch'' by considering the evolution of an elementary
particle when its energy approaches the Planck scale from below. In this sub-Planckian regime the increase of particle energy leads
to diminishing wave-length. However, when Planck energy is reached, a kind of ``phase transition'' takes place corresponding to an
\emph{increase} of wave-length with the energy. This non-standard behavior can be seen as the quantum translation of the relation 
between mass and radius of a classical BH. In other words, the quantum particle changes its nature by crossing the Planck barrier.  
Once it is given additional energy, it will
increase in size and eventually reach a semi-classical regime where the geometrical description  can be properly applied.\\
In the spirit of the above discussion, one may conclude that the quantum BH should be considered just another quantum particle, though with
a particular relation between its energy and size. 
\\
In recent papers \cite{Spallucci:2014kua,universe} we have made a first step towards the formulation of a truly quantum theory of BHs 
by starting with a simple one-dimensional model of a neutral BH. This toy-model has shown nice and simple quantization features, 
as well as, a natural limit towards a classical Schwarzschild BH for large principal quantum number.  \\
In this work we would like to extend the toy-model to a realistic three dimensional, charged BH, hopefully to be produced 
in the proton-proton collision at LHC.  To realize this project we are guided by the \emph{Holographic Principle } 
  \cite{Susskind:1994vu,Susskind:1998dq,Hooft:1999bw} claiming that
the whole dynamics of a quantum BH is the dynamics of its horizon .\\
  At first glance,  this statement is in clear contradiction with the purely
    geometric, and static, nature of a classical horizon. Thus, the first
    problem one encounters in trying to implement the Holographic
    Principle is how to introduce an intrinsic dynamics for
    the horizon. In the simplest case of a spherically symmetric
    BH, we are guided by the analogy with the two-body problem in the central
    potential where the relative dynamics can be described in terms of a 
    ``fictitious'' particle of reduced mass moving in a suitable one-dimensional
    \emph{effective potential}. Following the same line of reasoning, we started 
    by noting that the equation for the horizon(s) in the Reissner-Nordstr\"om geometry
    looks like the equation for the turning-points of a particle of energy $E=M$ moving between
    $r=r_-$ and $r=r_+$ where $r_\pm$ are the inner and outer horizons for a BH of mass $M$ and charge $Q$. Accordingly, 
    we propose to assign the horizon  an effective dynamics
    described by the motion of such a representative particle.  The motion of the particle
    in the interval $r_-\le r\le r_+$ corresponds to the  ``deformations'' of the horizon. 
    \\
In Section(\ref{class}) we  give an Hamiltonian formulation of the particle motion and solve the equation for the
orbits. Each orbit is characterized by a fixed value of the energy $E$($=M$ mass of the BH), the charge $Q$ ($=$ charge of the BH)
and angular momentum $L$. The motion of the particle is always bounded but the orbits are not always closed. 
\\
This particle-like model has the advantage to allow a straightforward quantization  leading to the corresponding quantum horizon model.\\
In Section(\ref{quantum}), we write and solve the horizon wave equation and determine the energy spectrum. As it can be expected from
the classical motion analysis, we find  \emph{discrete} energy levels depending from the radial quantum number $n$ and the orbital
quantum number $l$. Contrary to the classical description the BH mass, in the neutral case $Q=0$, cannot be arbitrarily small, but
is bounded from below by the ground-state energy $ E\simeq 1.22\times M_{Pl} $.\\
Finally, we find that in the classical limit $n>> 1$, the coordinate of the  peak of the probability density 
approaches the classical value for the horizon radius.\\
In the concluding Section (\ref{final}) we stress the modification our model introduces in the current picture of 
gravitational ``classicalization'' at the Planck scale.

\section{Particle analogue of a charged BH}
        \label{class}
The quantization of mechanical system, say a ``particle'', starts from a classical Hamiltonian encoding its motion.
On the other hand, a classical BH is defined as a particular solution of the Einstein equations. We give up such a starting
point in favor of a particle-like formulation translating in a mechanical language the key feature of a
BH which are summarized below:

            \begin{enumerate}
             \item BHs are intrinsically \emph{generally relativistic} objects, in the sense of strong gravitational fields.
              Thus, the equivalent particle model should start with 
               a relativistic-like dispersion relation for energy and momentum rather than a Newtonian one;
             \item the particle model must share the same spherical symmetry of the RNBH and the classical motion will
                   be described in terms of a radial and an angular degree of freedom;
             \item  the ``mass'' to be assigned to the horizon is the ADM mass;
             \item  The equation for the horizons,$r_\pm$, of a charged BH, looks like the mechanical equation
                    for the turning points of a particle with total energy $E=M$ in a suitable potential.
                    
            \begin{equation}
            M =\frac{r_\pm}{2\, G_N}\left(\, 1+\frac{Q^2G_N}{r^2_\pm}\,\right)\longleftrightarrow E= V(r_\pm)
             \end{equation}

                 \emph{This identification allows to map the problem of finding the horizons in a given metric into the
                 problem of determining the turning points for the bounded motion of a classical, relativistic, particle.}

              \end{enumerate}
            
         The above requirements are implemented through the following  Hamiltonian 

       \begin{equation}
	H\equiv \sqrt{\, \vec{p}^{\,\, 2} + m^2\left(\, r\,\right) }= 
        \sqrt{\, {p}^{\, 2}_r + \frac{p_\phi^{\,2}}{r^2} +\frac{r^2}{4\,G^2_N}\left(1+\frac{Q^2G_N}{r^2}\right)^2} \label{ham}
	\end{equation}
	Both the total energy and the angular momentum are constant of motion
       
         \begin{eqnarray}
          && \frac{\partial H}{\partial t}=0\longrightarrow H=const.\equiv E\ ,\\
          && \frac{\partial H}{\partial \phi}=0 \longrightarrow p_\phi= const.\equiv L
         \end{eqnarray}

          From the Hamilton equations we obtain 

          \begin{eqnarray}
           && \dot{r}^2= 1 - \frac{L^2}{E^2 r^2} - \frac{r^2}{4\,G^2_N E^2}\left(1+\frac{Q^2G_N}{r^2}\right)^2\ ,\label{h1}\\
           && \dot{\phi}^2= \frac{L^2}{E^2 r^4} \label{h2}
          \end{eqnarray}

            The parametric form of the solutions is:
            {\small
           \begin{eqnarray}
            && r(t)=\sqrt{2} G_NE \left[\, 1 - \frac{Q}{2G_N E^2} + \sqrt{1 -\frac{1}{G_N E^2}\left(\, Q^2 +\frac{L^2}{G_N E^2}\,\right)} 
             \cos\left(\, t/G_N E\,\right) \,\right]^{1/2} \label{param1}\\
            && \phi(t)=\frac{1}{\sqrt{1+\frac{Q^4}{4L^2}}}\arctan\left[\, \frac{L}{G_N E^2} 
            \frac{\sqrt{1+\frac{Q^4}{4L^2}}  \tan(t/2G_N E)  
            }{1 - \frac{Q}{2G_N E^2} + \sqrt{1 -\frac{1}{G_N E62}\left(\, Q^2 +\frac{L^2}{G_N E^2}\,\right)}}
            \,\right]\label{param2}
           \end{eqnarray}
            }
           A qualitative description of the motion can be obtained by writing equation (\ref{h1}) as the equation of motion
           for a particle in the \emph{effective potential}
          
          \begin{equation}
          \dot{r}^2= 1 - V_{eff}(r)^2 / E^2   \label{h1b}
          \end{equation}

          where 

         \begin{equation}
          V_{eff}(r)=\left[\, \frac{L^2}{ r^2} +\frac{r^2}{4\,G^2_N}\left(1+\frac{Q^2G_N}{r^2}\right)^2\,\right]^{1/2}
           \label{veff}         
\end{equation}

\begin{center}
\begin{figure}[h!]
 \includegraphics[height=10cm]{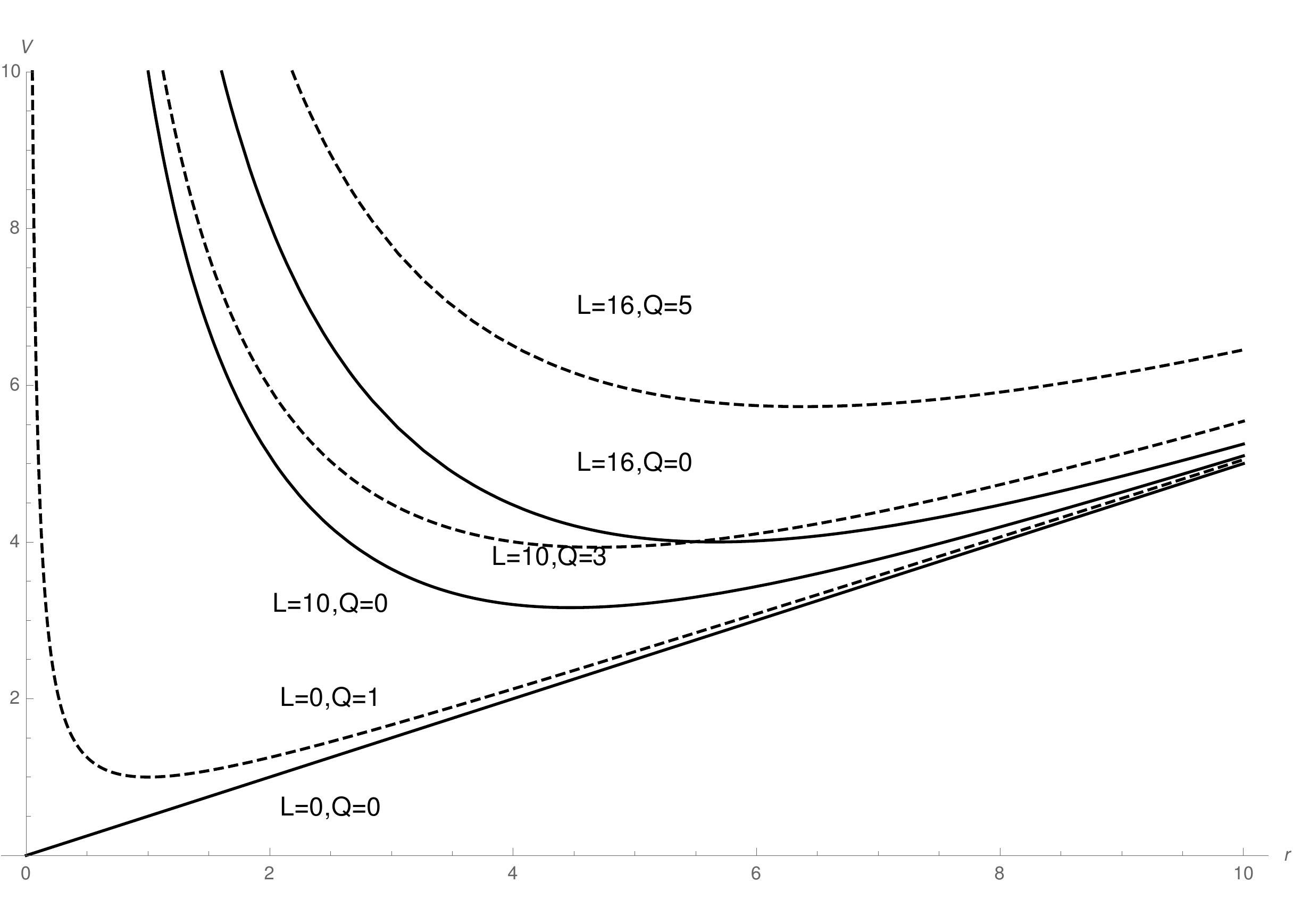}
\caption{Plot of the equation (\ref{veff}) for different values of $L$ and $Q$. }
\label{veff1}
\end{figure}
\end{center}

\begin{center}
\begin{figure}[h!]
 \includegraphics[height=10cm]{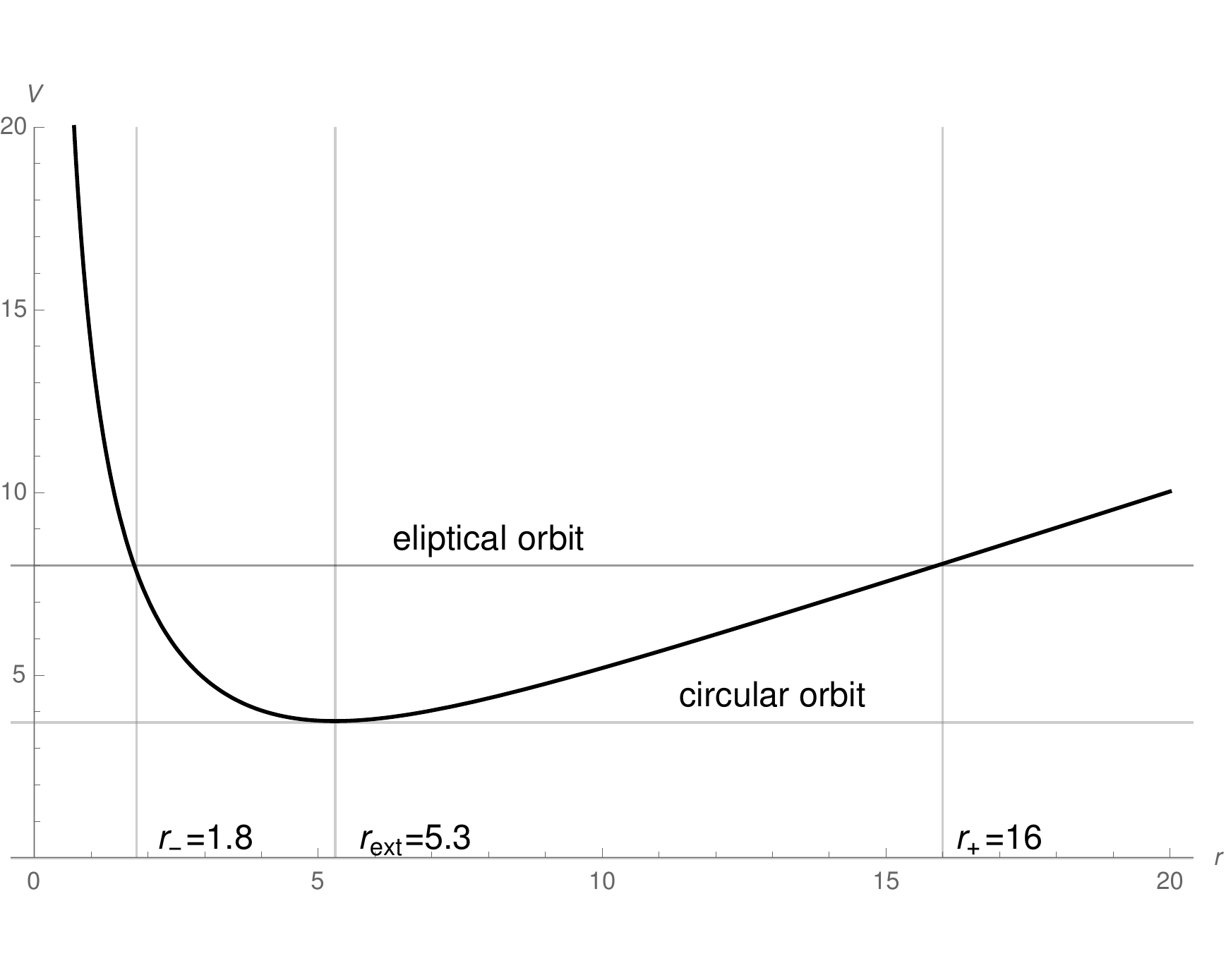}
\caption{Plot of $V_{eff}(r)$, with $L=14$, $Q=0$. $r_+=a$, $r_-=b$ are the turning-points corresponding to the maximum and and minimum
distance from the origin.
For $E=E_m $ the orbit degenerates into a circular orbit. }
\label{veff2}
\end{figure}
\end{center}
          The charge introduces an additional repulsive effect, at short distance, adding up to the centrifugal barrier. Instead, 
          at large distance 
          the charge-independent harmonic term is the leading one.\\
	  It follows that we have only \emph{bounded} orbits describing a bounded motion. This is in agreement with our purpose to model
           horizon vibrations around a stable equilibrium configuration in terms of the motion of a 
           representative
          ``particle''. In order to substantiate this analogy, let us 
           check, at first,  the correspondence between turning-points and horizon positions. \\
           
           \begin{equation}
             \frac{dV_{eff}(r)^2}{dr}=0\longrightarrow r_m^2 =2G_N L \sqrt{ 1 + \frac{Q^4}{4L^2} }
           \end{equation}
           The existence of a  minimum corresponds to a  stable circular orbits of radius $r_m$, or
           a static horizon of radius $r_+=r_m$
           
           \begin{equation}
            V_{eff}(r_m)= \frac{1}{2G_N }\left(\, Q^2 + \sqrt{Q^4 + 4L^2}\,\right)
           \end{equation}

           The energy of the particle on the circular orbit is given by

           \begin{equation}
            E^2_m= V_{eff}(r_m)  =\frac{1}{2G_N }\left(\, Q^2 + \sqrt{Q^4 + 4L^2}\,\right)
           \end{equation}
 
            and its angular frequency is
           \begin{equation}
           \dot{\phi}^2= \frac{L^2}{E^2_m r^4_m}=\frac{1}{2G_N}\sqrt{1 + \frac{Q^4 E^2_m}{4L^2}}
            \end{equation}

           For $E> E_m$ there are
          two turning points which are the solutions of the equation $\dot{r}=0$. By introducing the variable $ x\equiv r^2$, one gets the 
           algebraic quadratic equation

        \begin{equation}
         x^2 -2\left(\, 2G_N^2 E^2 - G_N Q^2\,\right)\, x + 4G_N^2 L^2 \left(\, 1 + \frac{Q^4}{4L^2}\,\right)=0 
        \end{equation}
  
           Thus,
                
         \begin{equation}
          r_\pm^2 =  \left(\, 2G_N^2 E^2 - G_N Q^2\,\right) \pm 2G_N E\sqrt{G_N^2 E^2 -Q^2G_N- L^2/E^2}\label{rpm}
         \end{equation}
        
          where

          \begin{equation}
           E^2 \ge \frac{Q^2}{2G_N}\left(\,  1 + \sqrt{ 1 + 4L^2/Q^4}     \,\right)\label{c1}
          \end{equation}

           For $L=0$ the condition (\ref{c1}) reduces to the condition $G_NE^2 \ge Q^2$ for the existence of the static RN horizons. 
           Furthermore, 
           the turning-points equation (\ref{rpm}) correctly gives the radius of both the inner (Cauchy) and outer (Killing) horizons.

           \begin{equation}
            r_\pm = G_N E \pm \sqrt{G_N^2 E^2 -Q^2G_N}
           \end{equation}

        From the Hamilton equations (\ref{h1}),(\ref{h2}) one obtains the orbit equation

        \begin{equation}
         \left(\, \frac{dr}{d\phi}\,\right)^2= 
         \frac{E^2 r^4}{L^2}\left[\,1 - \frac{L^2}{E^2 r^2} - \frac{r^2}{4\,G^2_N E^2}\left(1+\frac{Q^2G_N}{r^2}\right)^2 \,\right]
        \end{equation}

         which can be integrated:
         
         \begin{eqnarray}
          && r^2\left(\, \phi\,\right) = \frac{2L^2}{E^2}\left(\, 1 + \frac{Q^4}{4L^2}\,\right)\times\nonumber\\
              && \frac{1}{1-\frac{Q^2}{2G_N E^2} + 
         \sqrt{1-\frac{1}{G_N E^2}\left(\, Q^2 +\frac{L^2}{G_N E^2} \,\right) }
         \sin\left[\, 2\sqrt{ 1+\frac{Q^4}{4L^2} }\left(\, \phi-\phi_0\,\right) \,\right] }
         \label{orbite}
         \end{eqnarray}
          
       where $\phi_0$ is an arbitrary integration constant. The same solution can be obtained by eliminating time from
       equation (\ref{param1}),(\ref{param2}).\\
       The orbit equation (\ref{orbite}) can be conveniently re-written as
     
       \begin{equation}
           r^2\left(\, \phi\,\right) = \frac{2L^2\beta^2}{E^2} \frac{1}{1-\frac{Q^2}{2G_N E^2} - 
         \sqrt{1-\frac{1}{G_N E^2}\left(\, Q^2 +\frac{L^2}{G_N E^2} \,\right) }
         \cos\left[\, 2\beta\, \phi \,\right] }
         \label{orbite2}
         \end{equation}

where

       \begin{eqnarray}
        && \beta\equiv \sqrt{ 1+\frac{Q^4}{4L^2} }\ ,\\
        && \phi_0=\pi/4\beta
       \end{eqnarray}

To understand the property of the orbit,  let us consider the neutral BH $Q=0$ first.  This case describes the dynamics of the 
Schwarzschild horizon.

\subsection{Neutral orbits $Q=0$}

For $\beta =1$ the orbits simplify to 

        \begin{equation}
        r\left(\,\phi\,\right)=\frac{\sqrt{2}L}{E}\frac{1}{\left[\, 1-\sqrt{1-L^2/G^2E^4}\,\cos\left(\, 2\phi\right)\,\right]^{1/2}}
        \label{ell}     
\end{equation}

\begin{center}
\begin{figure}[h!]
 \includegraphics[height=10cm]{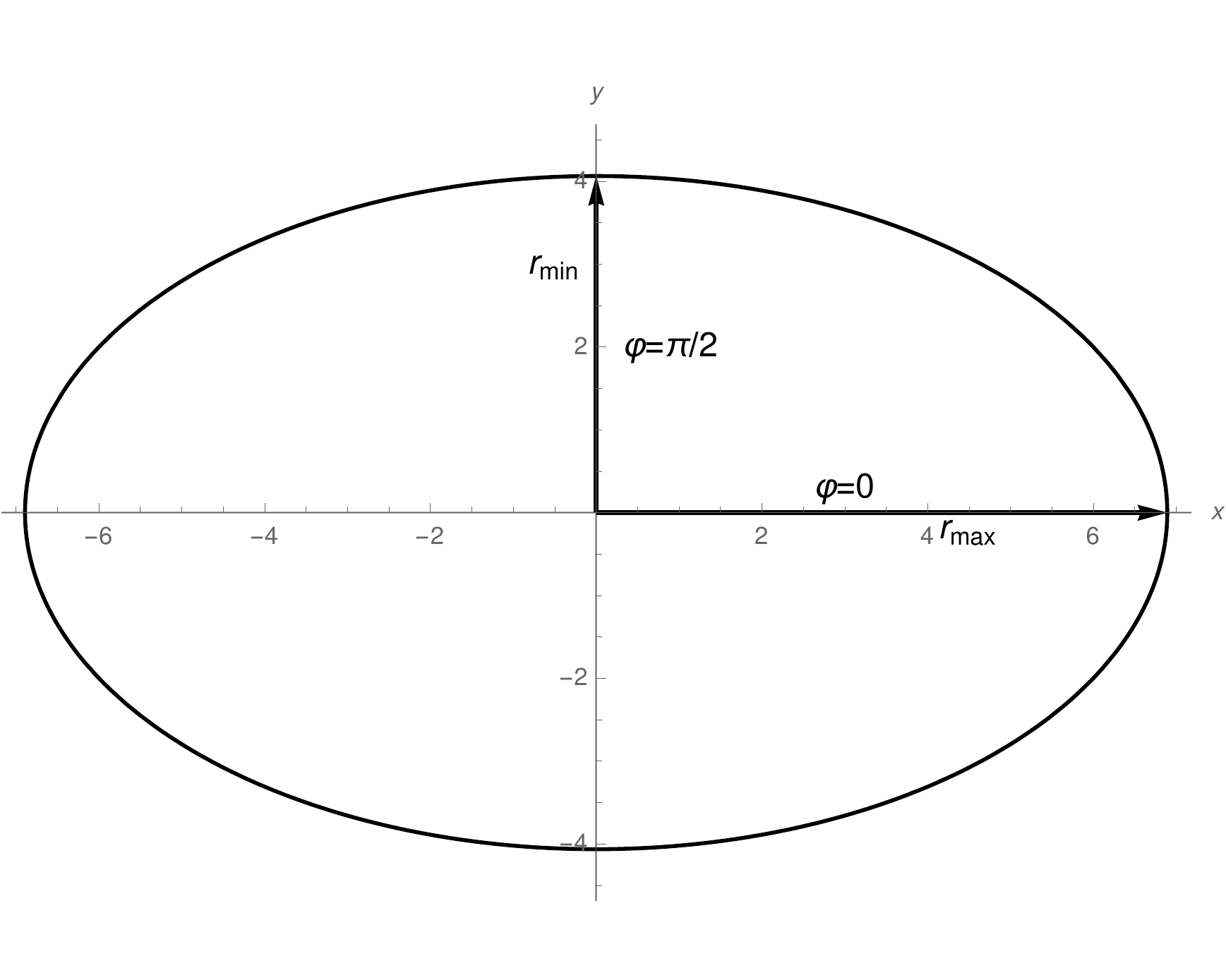}
\caption{Plot of the equation (\ref{ell}) in terms of the rescaled variables $r/\sqrt{G_N}$  with  $L=14$, $\sqrt{G_N} E=4$. }
\label{orbitaneutra}
\end{figure}
\end{center}
 Equation (\ref{ell}) describes  ellipses centered at the origin with major and minor semi-axis, $a$ and $b$ respectively, given by
\begin{eqnarray}
 && a=\sqrt{2} G_N E \sqrt{1+ \sqrt{1-L^2/G_N^2 E^4}}\ ,\\
 && b=\sqrt{2} G_N E \sqrt{1- \sqrt{1-L^2/G_N^2 E^4}}\ ,\\
 && L \le G_N E^2
\end{eqnarray}
This type of orbits correspond to a radially ``\emph{breathing}'' mode of the Schwarzschild horizon:
\begin{equation}
 \frac{\sqrt{2}L}{E}\frac{1}{\left[\, 1+\sqrt{1-L^2/G^2E^4}\,\right]^{1/2}}\le r(\phi)\le
\frac{\sqrt{2}L}{E}\frac{1}{\left[\, 1-\sqrt{1-L^2/G^2E^4}\,\right]^{1/2}}
\end{equation}

Two limits are of special interest. \\
For $L\to 0$ ellipses degenerates into a segment and the motion becomes e one-dimensional oscillation between the origin and the Schwartzschild
 radius $a= 2G_N E$, while $b= 0$.\\
The other limiting case is $L= G_N E^2$. In this case, the ellipse degenerate into a circle of radius $r= \sqrt{2} G_N E$ and 
the horizon ``freezes'' into a static configuration. $E=\sqrt{L/G_N}$ is the ground state energy corresponding to the stable 
minimum of the effective potential. \\
We recall that $r$ corresponds to the radius of the  BH. The existence of $r_{min}$ and $r_{max}$, for $L\ne 0$,
defines the range of radial vibrations of the Schwarzschild horizon.  To clarify the role of angular momentum we plot below
orbits for different $L$

\begin{center}
\begin{figure}[h!]
 \includegraphics[height=6cm]{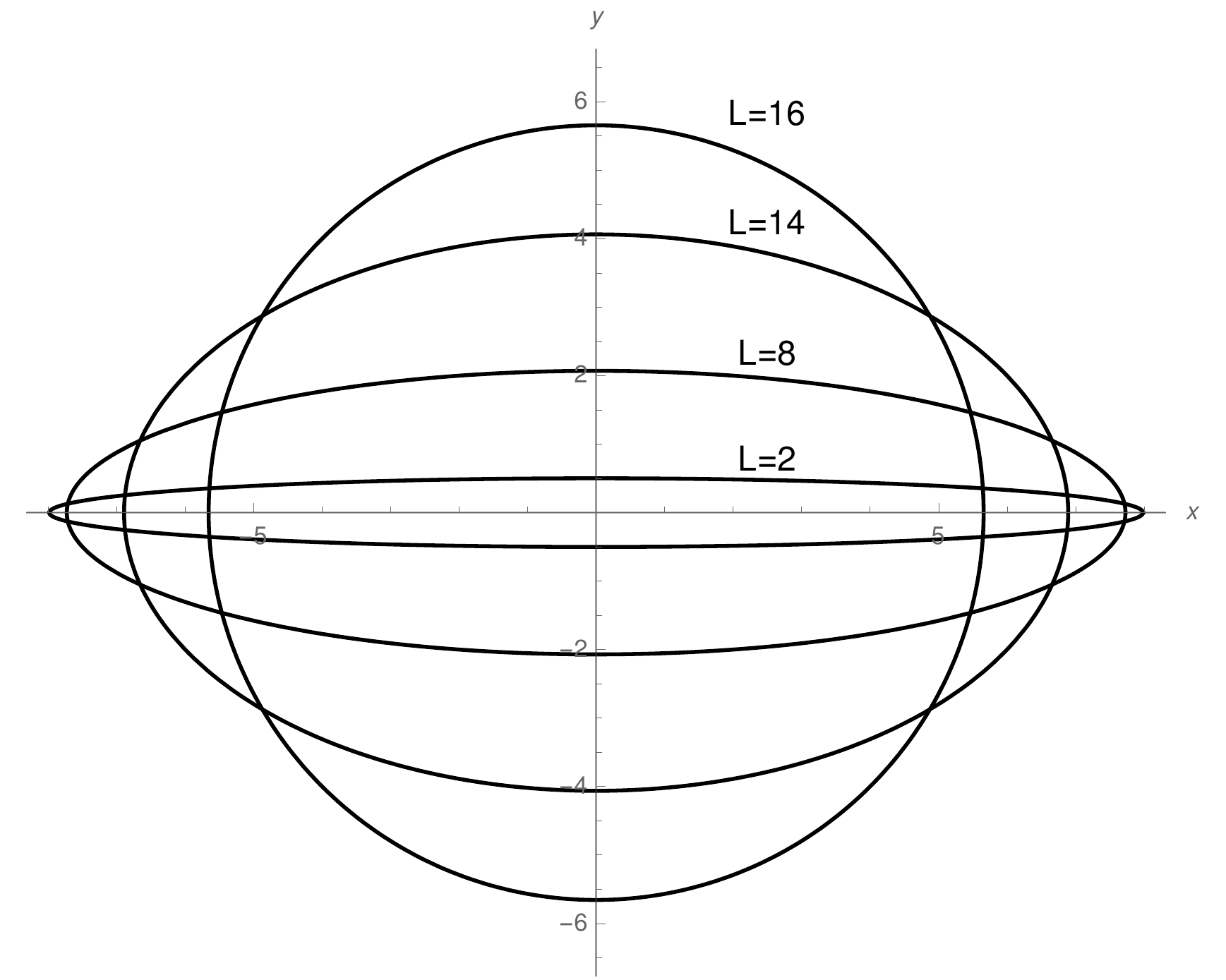}
\caption{Plot of the equation (\ref{ell}) for different values of $L$. $L=16$ is the limiting value corresponding to a circular orbit.}
\label{orbitaL}
\end{figure}
\end{center}
         
The figure (\ref{orbitaL}) clearly shows that there exist a maximum value of $L=G_NE^2$, for any given $E$, corresponding to 
the circular orbit.
Let us remark that, as it is expected, for $L=0$ $r_{max}=r(\phi=0) = 2G_N E$ is the Schwarzschild radius and
          $r_{min}=r(\phi=\pi/2)=0$. In the absence of angular momentum the whole problem collapses into a one-dimensional 
          harmonic motion.

\subsection{Charged orbits $Q\ne 0$}
       When $Q\ne 0$ the general solution of the orbit equation reads
 \begin{equation}
          r^2\left(\, \phi\,\right) = \frac{2L^2}{E^2}
              \frac{1 + Q^4/4L^2}{1-\frac{Q^2}{2G_N E^2} -
         \sqrt{1-\frac{1}{G_N E^2}\left(\, Q^2 +\frac{L^2}{G_N E^2} \,\right) }
         \cos\left[\, 2\beta\, \phi \,\right] }
         \end{equation}
describing a bounded motion of the particle around the origin. Again orbits are not always closed.

\subsection{Closed orbits}
Orbits are closed only if $\beta=n$, $n=2\ ,3\ , 4\, \dots$.\\

\begin{equation}
          r^2_{closed}\left(\, \phi\,\right) = \frac{2L^2}{E^2}
              \frac{n^2}{1-\frac{L\sqrt{n^2-1}}{G_N E^2} -
         \sqrt{1-\frac{1}{G_N E^2}\left(\, 2L\sqrt{n^2-1} +\frac{L^2}{G_N E^2} \,\right) }
         \cos\left[\, 2n\, \phi \,\right] }
         \end{equation}

with

\begin{equation}
 E^2\ge \frac{L}{G_N}\left(\, \sqrt{n^2-1} +n\,\right)
\end{equation}

\begin{center}
\begin{figure}[h!]
 \includegraphics[height=12cm]{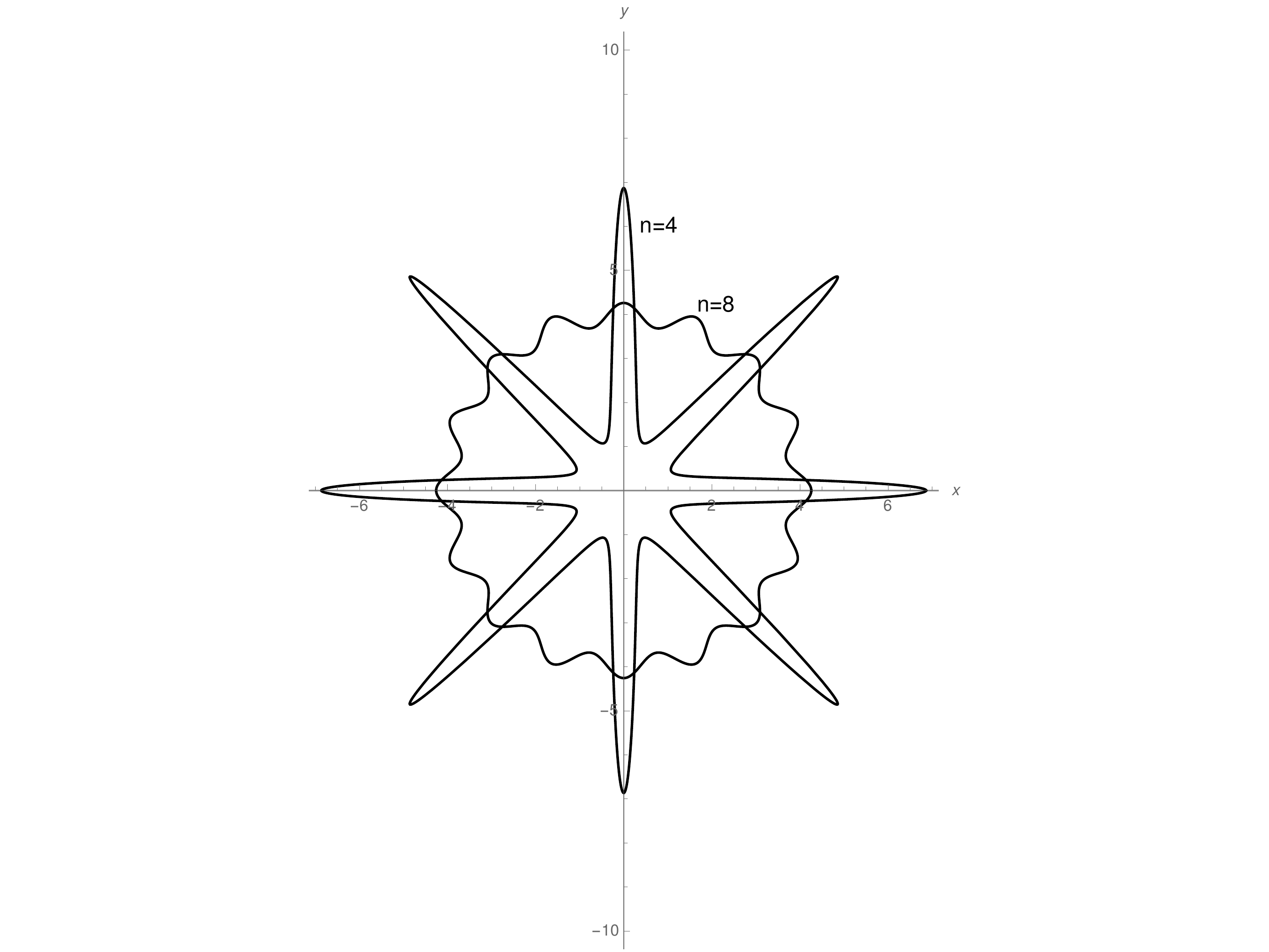}
\caption{Plot of two closed orbits with $n=4$ and $n=8$}
\label{orbitaneutra2}
\end{figure}
\end{center}

\subsection{Open orbits}

For $\beta \ne n$ orbits are \emph{open} and rotate by an angle $\Delta\phi =\pi/\beta$ every revolution Fig.(\ref{porb}).\\

\begin{equation}
          r^2_{open}\left(\, \phi\,\right) = \frac{2L^2}{E^2}
              \frac{\beta^2}{1-\frac{L\sqrt{\beta^2-1}}{G_N E^2} -
         \sqrt{1-\frac{1}{G_N E^2}\left(\, 2L\sqrt{\beta^2-1} +\frac{L^2}{G_N E^2} \,\right) }
         \cos\left[\, 2\beta\, \phi \,\right] }
         \end{equation}

Whatever is the value of $\beta$, we can compute the maximum and minum distance from the origin.

\begin{equation}
 \frac{dr^2}{d\phi}=0\longrightarrow \sin\left(\, 2\beta \phi\,\right)=0\longrightarrow \phi_k=k\frac{\pi}{2\beta}\le 2\pi
\end{equation}

with $k=0\ ,1\ ,2\ ,3\ ,\dots$.

\begin{equation}
 r^2\left(\,\phi_k\,\right)=r^2_k=\frac{2L^2\beta^2}{E^2}
\frac{1}{1-\frac{Q^2}{2G_N E^2} + \left(\,-1\,\right)^{k+1}\sqrt{1-\frac{1}{G_N E^2}\left(\, Q^2 +\frac{L^2}{G_N^2 E^2} \,\right)}}
\end{equation}
$k$ odd gives minimum distance $r_-$, and $k$ even gives maximum distance $r_+$. The limit $L\to 0$ is ``singular'' in the sense
that $\beta\to\infty$ and the orbit degenerates in a one-dimensional motion over the interval $r_- \le r \le r_+$:

\begin{equation}
 r^2\left(\,\phi_k\,\right)\to r_\pm^2 = 2G_N^2 E^2 - Q^2 G_N \pm 2G_N E \sqrt{G_NE^2 -Q^2}
\end{equation}
For vanishing angular momentum we recover spherical symmetry and the trajectory describes the oscillation of the
horizon between the inner and outer Reissner-Nordstrom radii:

\begin{equation}
 r_\pm = EG_N \pm \sqrt{E^2G_N^2 - Q^2 G_N}
\end{equation}

Finally, we notice that for

\begin{equation}
 2G_N E^2 = Q^2 \left[\, 1 + \sqrt{1 + \frac{4L^2}{Q^4}}\,\right] \label{e1}
\end{equation}

the orbit is $\phi$ independent, i.e. it is a circle
\begin{equation}
           r^2\left(\, \phi\,\right) = \frac{2L^2\beta^2}{E^2} \frac{1}{1-\frac{Q^2}{2G_N E^2}  }=2G_N \beta L
         \label{cerchio}
         \end{equation}

For $L\to$ (\ref{e1}) gives the extremality condition for the RN black hole $G_N E^2= Q^2$, and $r^2(\phi)\to G_N Q^2 = G_N^2E^2$.
Thus, the condition (\ref{e1}) represents a generalized \emph{extremality} condition in the presence of the angular momentum $L$.
\begin{center}
\begin{figure}[h!]
 \includegraphics[height=6cm]{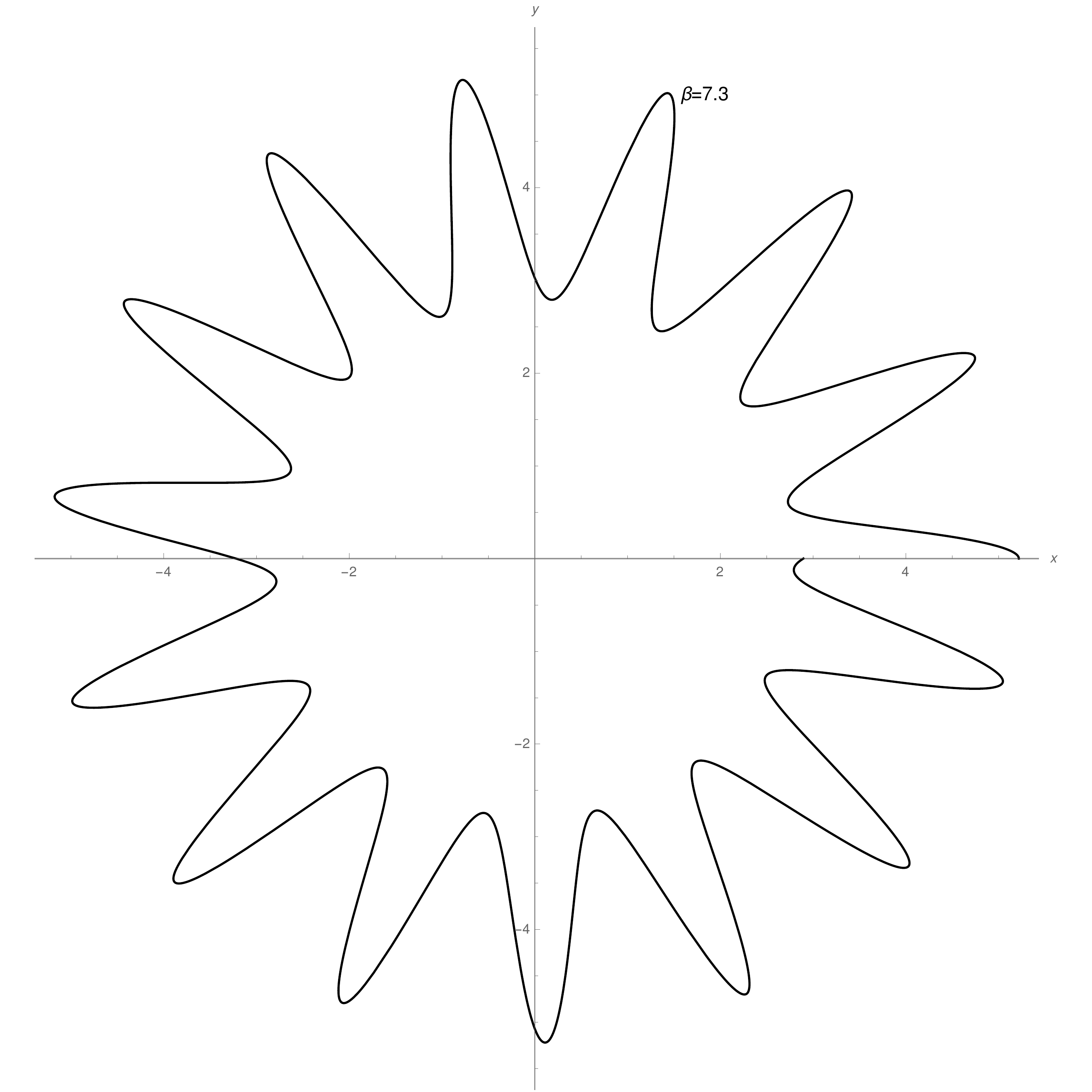}
\caption{Plots of an open orbit with $L=1$, $E=4/\sqrt{G_N}$,$\beta=7.3$ . }
\label{orbitechiuse}
\end{figure}
\end{center}

\begin{center}
\begin{figure}[h!]
 \includegraphics[height=8cm]{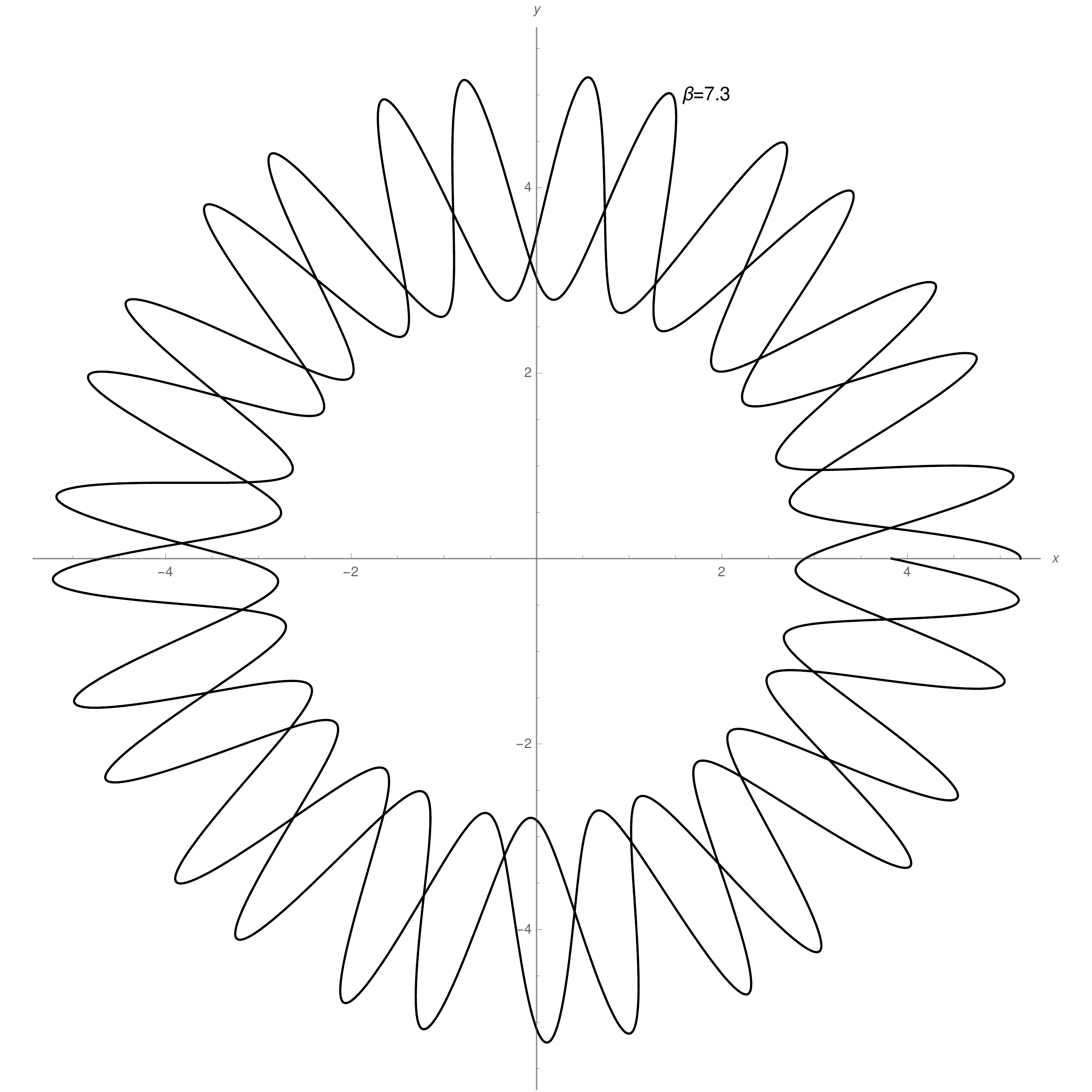}
\caption{Precession of the open orbit with $L=1$, $E=4/\sqrt{G_N}$,$\beta=7.3$ after two revolutions. }
\label{orbitechiuse2}
\end{figure}
\end{center}

\section{Quantum charged BH}
\label{quantum}

In this section we shall  quantize the classical model described previously. 
The quantization scheme contains the underlying idea 
to make the radius of the horizon(s) ``\emph{uncertain}'' and thus, unavoidably, described only in terms of a probability amplitude, 
or ``wave function''. From this perspective the horizon radius looses its classical geometrical  meaning.  
It acquires the role of wave-length of a Planckian BH. This description is motivated by the fact that in the
vicinity of the Planck scale the  wavelength of an ordinary quantum particle and the quantum mean radius of a Planckian BH 
merge and there is no distinction between the two.  Therefore, it is important to remark that a Planckian BH is
\emph{very different} from a (semi)classical one!  It is no more characterized by a one-way geometric boundary, but by a  wave-length 
which is an increasing function of the energy. Only far above the Planck scale, where the quantum fluctuations ``freeze-out'',  
one can resume the concept of classical horizon. \\ 
Our quantum description has a two-fold motivation:
\begin{itemize}
 \item it is generally accepted that the dynamics of a quantum gravitational system is completely  encoded in its boundary.
  This is the celebrated Holographic Principle which seems to find its natural realization in the quantum dynamics of a BH,
   where the ``boundary'' is the horizon itself. Already at the semi-classical level this principle is implied by the 
Bekenstein-Hawking    ``area law''.
\item As we have shown in the previous section, the classical horizon dynamics can be described in terms of a ``particle''
      moving in a suitable self-gravitational potential. Thus, it is straightforward to proceed by looking for the horizon wave function
      as the solution of a quantum wave equation for the corresponding classical particle studied before.
\end{itemize}

Starting from the classical Hamiltonian (\ref{ham}), following the standard quantization procedure, one obtains the corresponding 
wave equation a
%
%
\begin{eqnarray}
&& \left[\, \frac{1}{r^2}\frac{\partial}{\partial r}\left(\, r^2 \frac{\partial}{\partial r}\,\right) +\frac{1}{r^2 \sin\theta }
\frac{\partial}{\partial \theta}\sin\theta \frac{\partial}{\partial \theta} +\frac{1}{r^2\sin^2\theta}\frac{\partial^2}{\partial\phi^2}
\,\right]\Psi\left(\, r\ ,\theta\ ,\phi\,\right)\nonumber\\
&&+\left[\, E^2- \frac{r^2}{4\,G^2_N}\left(\, 1+\frac{Q^2G_N}{r^2}\,\right)^2\,\right]
\Psi\left(\, r\ ,\theta\ ,\phi\,\right)=0
\end{eqnarray}
The $O(3)$ symmetry of the problem allows to express the angular dependence of the wave function in terms of
spherical harmonics $Y_l^m\left(\, \theta\ ,\phi\,\right) $ as:

\begin{eqnarray}
 &&\Psi\left(\, r\ ,\theta\ ,\phi\,\right)=\psi(r)Y_l^m\left(\, \theta\ ,\phi\,\right)\ ,\\
 && l=0\ ,1\ ,2\ ,\dots \qquad -l\le m \le l  \label{hpsi}
\end{eqnarray}

Thus, the radial wave equation reads:

\begin{equation}
 \left[\, \frac{1}{r^2}\frac{\partial}{\partial r}\left(\, r^2 \frac{\partial \psi}{\partial r}\,\right) 
\,\right]\psi\left(\, r\,\right)
+\left[\, E^2- \frac{r^2}{4\,G^2_N}\left(\, 1+\frac{Q^2G_N}{r^2}\,\right)^2-\frac{l\left(\,l+1\,\right)}{r^2}\,\right]
\psi\left(\, r\,\right)=0
\end{equation}

The radial wave-function is given in terms of generalized Laguerre polynomials  $L_n^\alpha(x) $ as:

\begin{equation}
  \psi_{n}\left(\, r\,\right) = N_n\, \frac{r^{2s}}{\left(\,2G_N\,\right)^s}\, e^{-r^2/4G_N} \,L_n^{2s + 1/2}\left(\,r^2/2G_N\,\right)
\end{equation}

where 

\begin{equation}
 L_n^\alpha(x)\equiv \sum_{k=0}^n \frac{\Gamma(n+\alpha +1)}{\Gamma(n-k+1)\Gamma(\alpha +k +1)}\frac{(-x)^k}{k!}
\label{laguerre}
\end{equation}

and

\begin{equation}
4s\equiv \sqrt{Q^4 +\left(\, 2l+1\,\right)^2} -1
 \end{equation}

The normalization coefficient $N_n$ is recovered from the unitarity condition

\begin{equation}
 4\pi \int_0^\infty dr r^2 \vert \psi \vert^2=1\longrightarrow 
N_n= \frac{1}{2}\frac{\sqrt{n!}}{\sqrt{ \sqrt{2}\, \pi\, G_N^{3/2}  \Gamma\left(\, n+2s +3/2\,\right)}}
\end{equation}

As it is expected from the classical analysis of the particle motion, one obtains 
a discrete energy spectrum at the quantum level:

\begin{eqnarray}
 2G_N E_n^2 -Q^2&&=  4n +2 +\sqrt{ Q^4+ \left(2l+1\right)^2} \ ,\nonumber\\
            &&=  4\left(\, n + s\,\right) +3 \ ,
\qquad n=0\ ,1\ ,2\ ,\dots\label{spettro}
\end{eqnarray}

Equation (\ref{spettro})is a concrete and simple realization of the general conjecture that mass spectrum of a quantum BH 
should be discrete \cite{Dvali:2011nh,Calmet:2012fv}. 
Furthermore, the result shows that a quantum BH is significantly different from its classical counterpart. 
In fact, even in the neutral case, $Q=0$,  a stable, \emph{non-singular} ground state configuration with $n=0$ does exist.
The ground state energy is finite and close to the Planck energy

 \begin{equation}
 E_0 = \sqrt{\frac{3}{2}}\, M_P \approx 1.22\times M_{P}
\end{equation}

This is the \emph{lightest}, stable, BH physically admissible, and no physical process can decrease its mass below this lower bound. 
The true ground state of a quantum BH is free from all the pathologies of semi-classical, geometrical, BHs, e.g. singularities, 
thermodynamical instability, etc.\\
This is to be expected since all the semi-classical arguments loose their meaning at the truly quantum level.\\
Having acquired the notion that Plankian BHs are quite different objects from their classical ``cousins'', we would like to address
the question of how to consistently connect Planckian and semi-classical BHs. As usual, one assumes that the quantum
system approaches the semi-classical one in the ``large-$n$'' limit in which the energy spectrum becomes continuous.
Before doing so, let us first consider the radial density describing the probability of finding the particle at distance $r$ 
from the origin, define as $ p_n(r)\equiv 4\pi r^2 \vert \psi \vert ^2 $:

\begin{equation}
 p_n(x)=\frac{2\, n!}{\Gamma\left(\, n+2s +3/2\,\right)}\, x^{4s+2}\, e^{-x^2}\, \left(\, L_n^{2s +1/2}\left(\,x^2\,\right)\,\right)^2\ ,
\quad x\equiv r/\sqrt{2G_N}\label{prob}
\end{equation}

\begin{figure}[h!]
 \includegraphics[height=10cm]{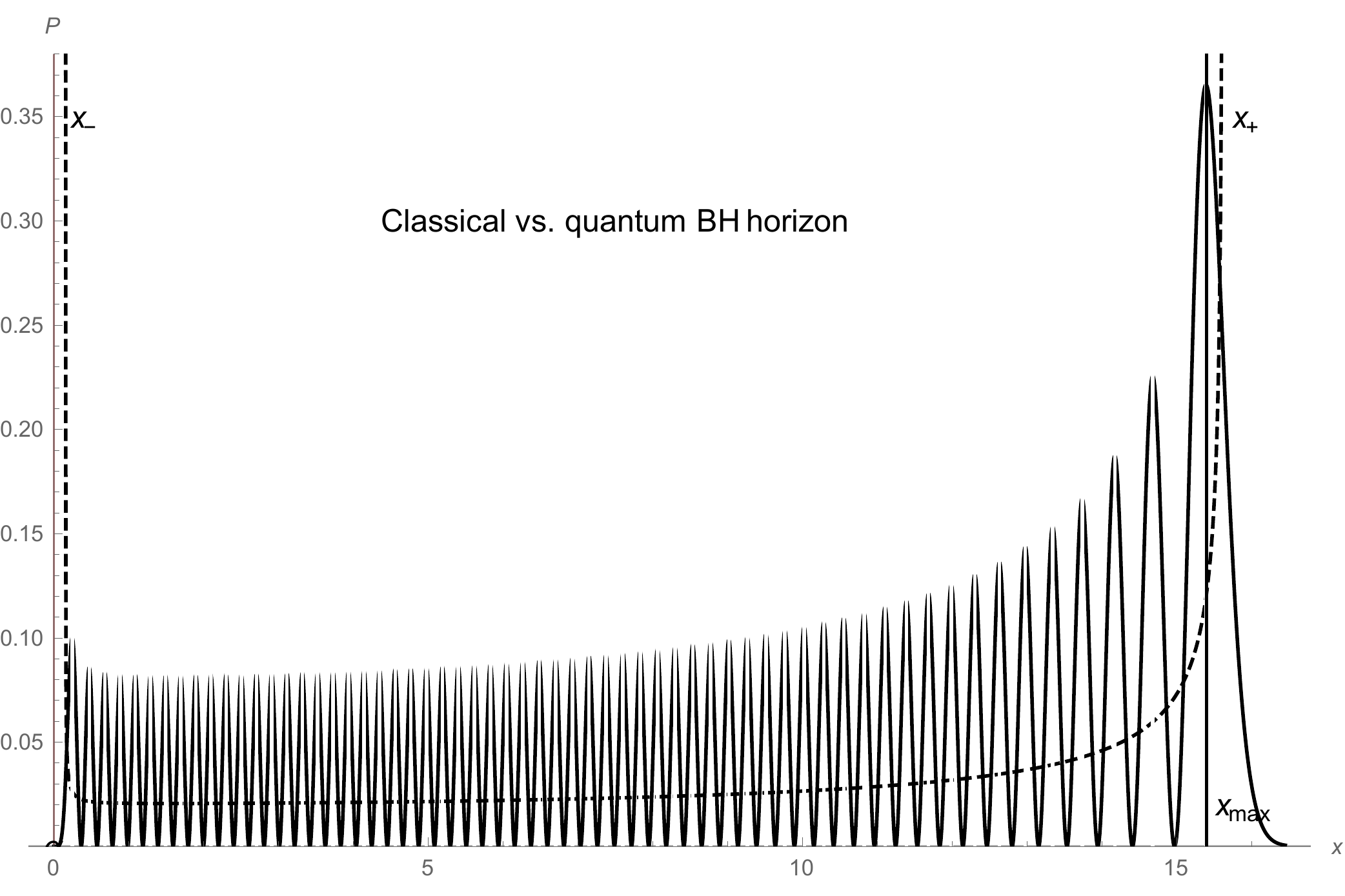}
\caption{Plot of the function $p_{n=60}(x)$, $s=1$ (continuous line) vs classical probability (dashed line). For large $n$ the
position of the first peak approaches $r_-$, while the last peak approaches $r_+$.}
\label{massimi}
\end{figure}

The local maxima in figure(\ref{massimi}) represent the most probable size of the Planckian BH.  
These maxima  are solutions of the equation

\begin{equation}
\left(\, 2s +1 -x^2 +4n\,\right)\, L_n^{2s+1/2}\left(\, x^2\,\right)-2\left(\, 2n +2s +1/2\,\right)\, 
L_{n-1}^{2s+1/2}\left(\, x^2\,\right)=0 \label{maxima}
\end{equation}

Equation (\ref{maxima}) cannot be solved analytically , but its large-$n$ limit can be evaluated as follows.
First, perform the division $L_{n}^{2s+1/2}/L_{n-1}^{2s+1/2}$, and then write

\begin{equation}
 L_n^{2s+1/2}\left(\,x^2\,\right)=P_2\left(\, x^2\,\right)\, L_{n-1}^{2s+1/2}\left(\, x^2\,\right)+Q_{n-2}\left(\, x^2\,\right)\label{ratio}
\end{equation}

where, 

\begin{eqnarray}
&& P_2 =\frac{a_n}{b_{n-1}}\left(\, x^2-2n-2s+1/2\, \right)\ ,\\
&& Q_{n-2}\left(\, x^2,\right)=c_{n-2}x^{2n-4}+\cdots\nonumber\\
&&\>\>\>\>\> \>\>\>\>\>\>\>\>\>\>\>\>\>\>\>\>\>\>=-\left(\, n-1\,\right)\left(\, n+2s-1/2\,\right)\, a_n\,x^{2n-4}+\cdots
\end{eqnarray}

By inserting equation (\ref{ratio}) in equation (\ref{maxima}) and by keeping terms up order $x^{2n-2}$, the equation
for maxima turns into

\begin{equation}
\left[\,x^2-2(n+s)-1\,\right]\,\left[\, x^2-2(n+s)+1/2\,\right]+\left[\, 2n+4s+1\,\right]\frac{b_{n-1}}{a_n}=(n-1)(n+2s-1/2)
\label{max1}
\end{equation}

where the coefficients of the of $L_n^{2s+1/2}$ and $L_{n-1}^{2s+1/2}$ from (\ref{laguerre}) are given by

\begin{eqnarray}
&& a_n=\frac{(-1)^n}{n!}\ ,\\
&& b_{n-1}=\frac{(-1)^{n-1}}{(n-1)!}
\end{eqnarray}

Equation (\ref{max1}), for large $n$ reduces to
 \begin{align*}
& 3n\left(n+2s\right)= \left(x^2-2\left(n+s\right)\right)^2\\
& x^2= 2\left(n+s\right)+\sqrt{3 n\left(n+2s\right)}\underbrace{\approx}_{s<< n}2\left(n+s\right)+\sqrt{3}n\left(1+s/n\right)
+\cdots\\
& x^2= (2+\sqrt{3})\left(n+s\right)=3,73\left(n+s\right)
\end{align*}

Thus, one finds the absolute maximum to be
\begin{equation}
 x^2=3.73\times\left(\, n+s\,\right)
\end{equation}

while, the classical radius of the horizon, for $E >> Q/\sqrt{G_N}$, is obtained by expressing (\ref{rpm}) in terms of $s$ 
and (\ref{spettro})

\begin{equation}
  \frac{r_+^2}{2G_N}\simeq 2G_NE^2 -Q^2 \simeq 4n +\sqrt{1+Q^4} \simeq 4\left(\, n+s\,\right)
 \end{equation}
 
which leads to

\begin{equation}
 x^2_+=4\,\left(\, n+s\,\right) \label{class2}
\end{equation}
 
Thus, we find that most probable value of $r$ approaches the horizon radius $r_+$ for $E>> M_p$, 
restoring the (semi)classical picture of BH.

\section{Discussion and future perspectives}
	\label{final}
In this closing section we would like to answer a couple of possible questions about our non geometric approach
to quantum BHs.\\ 
First of all, why should one use a single particle-like formulation?\\
Before answering this question one needs to explain what does it mean ``to quantize a BH''. Naively, one could 
say think to look for the amplitude to find the BH somewhere in space at a given instant of time. This is not the
case because we are not interested in the \emph{global} quantum dynamics of the object, but rather to its ``\emph{internal}''
dynamics. At this point we face the problem to define what is the BH internal dynamics. In this respect the 
the Holographic Principle provide the road map. The internal dynamics is nothing but the horizon dynamics, but 
General Relativity does not provide any dynamics to the BH horizon which is a purely geometrical boundary. At the quantum,
level one expects the radius and the shape of the horizon to become \emph{uncertain}. Near the Planck scale the mean value
of the horizon radius $<r_+>$ becomes comparable, or even smaller, than the the uncertainty $\Delta r_+$ and the very concept
of geometrical description of the  horizon become meaningless.
Thus, the first step towards a quantum BH is to move away from the safe land of General Relativity towards an
uncharted territory. \\
In the  case of a spherically symmetric BH, we are guided by the analogy with the two-body problem in the central
    potential where the internal dynamics of the system can be described in terms of a 
    ``fictitious'' particle of reduced mass moving in a suitable one-dimensional
    \emph{effective potential}. Following the same line of reasoning, we started 
    by noting that the equation for the horizon in the Schwartzschild metric
    looks like the equation for the turning-points of a particle of energy $E$ moving between
    $r=r_-$ and $r=r_+$. Accordingly, we propose to assign the horizon an effective dynamics
    described by the motion of such representative particle.  The motion of the particle
    in the interval $r_-\le r \le r_+$ corresponds to the vibrational modes of the  horizon.
    Thus, we conclude that our particle-like approach provides a simple and effective implementation of the Holographic
    Principle.\\

    The second important question to answer is how does a geometric picture of the horizon emerge from the quantum description.\\
    The classical limit is, perhaps, the most delicate feature of any quantum
    theory. Nevertheless, in our case, the answer should be pretty clear. The wave function (\ref{hpsi})
    is the probability amplitude to find the BH with an horizon of radius $r_+$. As the probability 
    density (\ref{prob}) and the plot in Fig.(\ref{massimi}) show, there are many possible values of the 
    horizon radius for a given energy level $E_n$, but there is a single highest peak of the probability
    density.  For $ E_n>> M_P$, the peak approaches the classical classical radius $r=r_+$. This behavior
    is clearly shown in Eq.(\ref{class}). Thus,  the geometrical
    picture of the horizon is recovered in the sense that the most probable value of the horizon radius reduces to
    the classical value provided by General Relativity in a far trans-Planck regime.\\
     
    Having clarified the two main points above, let us conclude this paper with a brief comment about 
    elementary particles and  Planckian BHs.\\   
The underlying idea that motivated this paper is the generally accepted view that, at the Planck scale,
a kind of ``~transition~'' between particles and micro-BHs takes 
place \cite{Markov:1967lha,Markov:1972sc,Aurilia:2013mca}. In detail, an elementary particle, in the
sub-Planckian regime, has its wavelength inversely proportional to its energy, but when it crosses the ``Planck energy
barrier'' this relation suddenly changes into a direct proportionality. This new relation between energy and
wavelength is associated with the appearance of a micro-BH because this kind of relation is characteristic of the BH 
horizon radius and its mass. In recent so-called UV self-complete quantum gravity program, this transition has been 
called ``classicalization'' \cite{Dvali:2011th,Dvali:2012mx} 
in the sense that a quantum particle turns at once in a \emph{classical}, even if microscopic,
BH. Although we are in agreement with this general picture, we presented in this letter its refined version, in the sense
that, in our view,  classicalization does not take place immediately, but much above the Planck energy barrier.
The intermediate region, immediately above the Planck scale, is dominated by pure quantum objects which have all
the characteristics of a quantum particle, except for the relation between its wavelength and energy. These objects could
be tentatively called quantum Planckian BHs bearing in mind that they are very different from the their (semi)classical
counterparts. However, they deserve the name ``black holes'' because we have shown that in the high energy limit they
grow into  (semi)classical BHs as we know them. The main difference between these two families bearing the same name 
``black holes'' resides in the fact that the Planckian BHs have no horizon in the classical sense and no geometric
interpretation. They behave and interact as ordinary quantum particles, and even if there will be no available energy to produce
them in high energy experiments, they should be taken into account as virtual intermediate states. From this point of view,
it is possible to expect to measure their indirect effects in particle collisions even at energy much below the Planck scale.
The most promising scenario for this effects to be seen is within 
 \emph{large extra-dimension} models \cite{Gabadadze:2003ii}, in which the Planck scale can be lowered not too far from 
the TeV scale.

\end{document}